\documentclass[conference]{IEEEtran}
\IEEEoverridecommandlockouts
\usepackage{cite}
\usepackage{booktabs}
\usepackage{url}
\usepackage{subcaption}
\usepackage{amsmath,amssymb,amsfonts}
\usepackage{algorithmic}
\usepackage[ruled,vlined]{algorithm2e}
\usepackage{graphicx}
\usepackage{textcomp}
\usepackage[normalem]{ulem}
\usepackage{multirow}
\usepackage{rotating}
\usepackage{tabularray}
\usepackage{xcolor}
{ \newcommand{\mynote}[2]{
      \fbox{\bfseries\sffamily\scriptsize#1}
        {\small\#{{{#2}\bf }}\#}}}
        { \newcommand{\mynote}[2]{}}

\def\BibTeX{{\rm B\kern-.05em{\sc i\kern-.025em b}\kern-.08em
    T\kern-.1667em\lower.7ex\hbox{E}\kern-.125emX}}
\begin{document}

\title{EUREKHA: Enhancing User Representation for Key Hackers Identification in Underground Forums\\

}
\author{
\IEEEauthorblockN{Abdoul Nasser Hassane Amadou, Anas Motii, Saida Elouardi, EL Houcine Bergou}
\IEEEauthorblockA{\textit{College of Computing} \\
\textit{University Mohammed VI Polytechnic}\\
Benguerir, Morocco \\
{\{abdoul.amadou, anas.motii, saida.elouardi, elhoucine.bergou\}}@um6p.ma}
}



\maketitle

\begin{abstract}

Underground forums serve as hubs for cybercriminal activities, offering a space for anonymity and evasion of conventional online oversight. In these hidden communities, malicious actors collaborate to exchange illicit knowledge, tools, and tactics, driving a range of cyber threats—from hacking techniques to the sale of stolen data, malware, and zero-day exploits. Identifying the key instigators (i.e., key hackers), behind these operations is essential but remains a complex challenge. This paper presents a novel method called \textit{EUREKHA} (Enhancing User Representation for Key Hacker Identification in Underground Forums), designed to identify these key hackers by modeling each user as a textual sequence. This sequence is processed through a large language model (LLM) for domain-specific adaptation, with LLMs acting as feature extractors. These extracted features are then fed into a Graph Neural Network (GNN) to model user structural relationships, significantly improving identification accuracy.
Furthermore, we employ BERTopic (Bidirectional Encoder Representations from Transformers Topic Modeling) to extract personalized topics from user-generated content, enabling multiple textual representations per user and optimizing the selection of the most representative sequence. Our study demonstrates that fine-tuned LLMs outperform state-of-the-art methods in identifying key hackers. Additionally, when combined with GNNs, our model achieves significant improvements, resulting in approximately 6\% and 10\% increases in accuracy and F1-score, respectively, over existing methods. \textit{EUREKHA} was tested on the \textit{Hack-Forums}\footnote{https://hackforums.net/}  dataset, and we provide open-source access to our code\footnote{\url{https://github.com/jumbo110/EUREKHA}}.

\end{abstract}


\section{Introduction}
To stay ahead of the ever-changing threat landscape, embracing proactive security measures such as Cyber Threat Intelligence (CTI) is essential. CTI involves gathering, analyzing, and utilizing data on security threats, threat actors, malware, vulnerabilities, exploits, and indicators of compromise (IoCs). An effective CTI must deliver relevant and precise information, identify credible threats, and offer actionable insights for threat mitigation.
Through rigorous examination, recent research highlights the increasing importance of underground markets and communities~\cite{du2019identifying,faisal2019expert}, delving into their socio-economic and geographical ramifications.

These underground forums frequented by hackers serve as valuable sources of CTI. A study~\cite{CASCAVILLA2021102258} reveals that the intelligence from hacker forums is underutilized in addressing cyber threats. This reluctance can be attributed to the overwhelming volume of unprocessed data in these forums, which poses a significant processing challenge. Many of these data are irrelevant or repetitive, making the manual analysis of hacker forum data a laborious, resource-intensive, and error-prone process.
We define a "key hacker" as an influential user involved in cybercriminal activities, including distributing malware, offering ready-made denial-of-service attack tools, or using these tools to launch attacks on others. Identifying key hackers in underground forums is crucial; by pinpointing them, security teams can efficiently extract valuable CTI while ensuring the authenticity of the extracted content and its sources, all within a time-efficient framework.

This research is motivated by the following factors: (1) the critical importance of such information for effective incident response and detection within organizations, (2) the substantial volume of data originating from underground forums, and (3) the need to optimize security teams resources by focusing on key hackers, whose posts may provide valuable insights, including details on malware, techniques, and zero-day vulnerabilities.

Despite the strides made by previous studies~\cite{shafer2020, Pannu2019} in automating CTI extraction through the semantic interpretation of forum data, the persistent challenge remains in validating profiles before CTI extraction. 
In~\cite{Chen2021}, researchers developed a system that employs behavioral analysis and account recognition techniques via social networks and semantic analyses of users to identify cybercriminals. Moreover, the authors in~\cite{Peersman2021} used centrality measures to categorize user types, although their reliance on outdated data highlights the need for real-time data analysis for effective CTI. In \cite{zhang2019key}, the authors introduced an attributed heterogeneous information network designed to identify key hackers using a Graph Convolutional Network (GCN), achieving an accuracy rate of approximately 90\%. Other research efforts have tackled similar challenges in various domains, including the dark web \cite{Arnold2019, cabrero2021methodology, akyazi2021measuring}, organizational security \cite{cherqi2023}, public exploits \cite{Du2023}, and deceptive CTI \cite{Ranade2021}.

Recent hacker identification methods using GNNs~\cite{zhang2019key,zhang2018kadetector} have demonstrated potential but are limited by the underutilization of user attributes and lengthy training times due to shallow text embedding techniques, such as skip-gram\cite{mikolov2013efficient} and bag-of-words (BoW)\cite{harris1985distributional}. Additionally, reliance on embedding techniques like Doc2Vec\cite{lau2016empirical} and Word2Vec~\cite{johnson2024detailed} struggles to capture the full complexity of users' semantic features, especially for those with extensive posting histories. This leads to biased and low-quality embeddings due to the dilution of attribute information during node feature extraction. In contrast, multi-layer LLM models provide a more sophisticated and nuanced approach to text representation. Recently, the authors in~\cite{10.1145/3678890.3678930}, evaluated the efficiency of hacker forums as a source of Cyber Threat Intelligence, using fine-tuned BERT-based models to conduct the analysis.

We introduce a novel approach, Enhancing User Representation for Key Hackers Identification in Underground Forums (\textit{EUREKHA}), which leverages LLM for domain adaptation and feature extraction. We first convert the metadata, threads, and replies of each user into a unified textual sequence for representation. These sequences are then submitted to the LLM model, where we perform fine-tuning for domain adaptation to help the model acquire relevant domain knowledge. However, processing extensive user interaction data, such as years' worth of activity, is challenging due to the complexity and noise in lengthy user sequences. The variability in user content over time further complicates pattern recognition. To address this issue, we use BERTopic~\cite{grootendorst2022bertopic} to condense large posting histories into concise topics, reducing computational demands while preserving key information. By enhancing user profiles with a deeper semantic understanding, we enable GNNs to effectively combine textual attributes with nuanced user characteristics and graph topology. This integration creates a comprehensive source of information that greatly enhances key hacker identification. We summarize our contributions as follows:
\begin{itemize}

    \item We utilize BERTopic to extract personalized topics from user-generated content, summarizing extensive user history into relevant patterns and enabling the creation of four distinct user sequence representations. We evaluate these representations to identify the most effective one for key hacker identification.
    \item To the best of our knowledge, we are the first to explore LLM-based methods for key hacker identification. We fine-tune several LLM models for this task and find that they outperform traditional GNN-based methods. Our added value includes extracting features with LLMs instead of using classical embedding techniques such as Word2Vec and Doc2Vec.

    \item Our experiments show that \textit{EUREKHA} achieves state-of-the-art performance, outperforming existing methods by approximately 6\% in accuracy and 10\% in F1-score.

\end{itemize}

The remainder of the paper is organized as follows: Section~\ref{relatedwork} provides a review of existing literature. Section~\ref{approach} provides a brief overview of the \textit{EUREKHA} framework. Section~\ref{process} presents the proposed methodology in more detail, including experimental datasets, user representation, topic modeling of content, the fine-tuning of LLMs, and GNN training. Section~\ref{experiment} describes the experimental setup and evaluates the performance of different LLM and GNN models, comparing our results with recent work to demonstrate that we achieve the highest accuracy for key hacker identification. Section~\ref{limitations} discusses the limitations and suggests potential directions for future research. Finally, Section~\ref{conclusion} summarizes our contributions and concludes the paper.

\section{Related Work}\label{relatedwork}
Table~\ref{related-works} provides a comprehensive summary of related works focused on identifying key hackers in underground forums. Several methods have been proposed, which can be divided into three main categories: content-based methods, centrality-based methods, and graph neural network-based methods.
\subsubsection{\textbf{Content-based methods}}
The data generated by users of underground forums is substantial and includes a variety of elements, such as threads, posts, comments, and uploaded attachments. A content-based method involves mining this data and constructing user evaluation metrics to identify key hackers. The most common evaluation metrics are activity level and content quality. In~\cite{content_marin}, the authors analyzed content features, seniority features, and social network features within the context of underground forums. An optimization meta-heuristic was employed to identify key hackers, and a systematic reputation-based method was proposed to validate the findings. Fang et al.~\cite{content_fang} developed a framework incorporating topic models to extract popular topics, track topic evolution, and identify key hackers along with their specialties. In~\cite{content_zhang}, the authors investigated the transfer of knowledge among user posts in underground forums and classified users into four categories: expert, casual, learning, and novice hackers. Expert hackers are knowledgeable and respected members of the community who increasingly serve as knowledge providers. While content-based analysis provides comprehensive metrics reflecting user influence, it is also more complex, necessitating professional involvement and verification in the selection of evaluation metrics. 
\subsubsection{\textbf{Centrality-based methods}}
Centrality measures are commonly used to identify influential hackers based on social network structures. These studies often begin by using content filtering techniques, such as Latent Dirichlet Allocation (LDA)~\cite{lda}, to categorize content based on topics that align with specific skill levels and to filter out users who do not have a certain level of expertise. After filtering, social network structures are built by analyzing user interactions, followed by the application of centrality measures to identify influential hackers. In~\cite{samtani2016using}, the authors identify key hackers in a large English hacker forum who distribute keyloggers using only degree and betweenness centrality measures within a social network. Similarly, in~\cite{samtani2017exploring}, the authors predicted key hackers by analyzing various malicious hacker tools, relying on these same centrality measures. Recently, Huang et al.~\cite{huang2021hackerrank} introduced HackerRank, a tool designed to analyze and identify key hackers in underground forums. The tool uses LDA to filter out users with low hacking skills and then uses PageRank to highlight highly skilled hackers.
In~\cite{paracha2023sus}, the authors introduce an AI-driven framework (INSPECT) to identify key hackers in underground forums based on five centrality measures (degree, betweenness, closeness, eigenvector, and PageRank).

However, these approaches have significant limitations. Content filtering methods, such as LDA, can oversimplify user categorizations and miss nuanced skill indicators. Traditional LDA models may fail to capture these contextual differences. Additionally, while centrality-based methods focus on network structure, they overlook important user characteristics, such as trust, reputation, and prestige, which are crucial in underground forums. To overcome these challenges, emerging trends such as Graph Neural Networks (GNNs)~\cite{zhou2020graph} offer promising solutions.

\subsubsection{\textbf{GNN-based methods}}
Graph Neural Networks (GNNs) offer a robust alternative for detecting influential hackers, surpassing traditional centrality-based methods. GNNs excel at capturing complex, nonlinear relationships in the data. They can integrate user features, including behavioral patterns, interaction histories, and contextual attributes such as trust, reputation, and prestige. In~\cite{xu2022hghan}, the authors used a heterogeneous graph attention network (HGHAN)~\cite{wang2019heterogeneous} to develop a system for identifying hacker groups. In~\cite{zhang2018kadetector}, the authors use heterogeneous graph representation and meta-path~\cite{huang2022able} approaches to establish user relationships on \textit{Hack-Forums}. They introduced ActorHin2vec to learn low-dimensional user representations within the Heterogeneous Information Network (HIN) framework and identify key hackers. However, this approach does not consider crucial user attributes such as profiles, post content, and threads, which are essential for accurately representing users. To address this limitation, the authors of~\cite{zhang2019key} propose an Attributed Heterogeneous Information Network (AHIN) as a solution. This method integrates meta-paths to assess user relationships and introduces Player2Vec for efficient node representation in identifying key hackers. By incorporating user attributes into their model, they improve the accuracy and performance of user characterization in complex networks compared to ActorHin2vec.

Despite the performance of GNNs in learning user representations in underground forums, they are limited by shallow text embeddings that fail to capture complex user attributes, especially for users with extensive posting histories, leading to biased and low-quality embeddings due to the dilution of attribute information during node feature extraction. Our approach uses LLMs for feature extraction, generating enriched user representations as features for GNNs. This hybrid method enhances user representations, significantly improving key hacker identification in underground forums.
However, processing extensive user interaction data, such as years' worth of activity, is challenging due to the inherent complexity and noise in lengthy user sequences. To address this challenge, we employ BERTopic~\cite{grootendorst2022bertopic} to extract personalized topics from user-generated content, thereby reducing data complexity and enhancing LLM training for a deeper understanding of user profiles in underground forums.

\begin{table*}
\centering
\caption{Existing works on identifying key hackers in underground forums.}
\label{related-works}
\begin{tblr}{
  cell{1}{1} = {r=2}{},
  cell{1}{2} = {c=3}{c},
  cell{1}{5} = {r=2}{c},
  cell{1}{6} = {r=2}{},
  cell{3}{6} = {r=3}{},
  cell{6}{6} = {r=3}{},
  hline{1,12} = {-}{},
  hline{2} = {2-4}{},
  hline{3} = {2-6}{},
}
\textbf{Reference}         & \textbf{Methods} &                     &              & \textbf{Summary}                                                                                                       & Limitations                                                                                                                                                                                                  \\
                           & \textbf{Content} & \textbf{Centrality} & \textbf{GNN} &                                                                                                                        &                                                                                                                                                                                                              \\
\cite{content_marin}       & \checkmark       &                     &              & {They used a meta-heuristic to analyze \\content and seniority to identify key hackers.}                               & {While content-based analysis offers \\comprehensive metrics that reflect \\user influence, it is more complex and\\requires professional involvement and \\verification in selecting evaluation \\metrics.} \\
\cite{content_fang}        & \checkmark       &                     &              & {They developed a framework with topic models\\~to identify key hackers and their specialties.}                        &                                                                                                                                                                                                              \\
\cite{content_zhang}       & \checkmark       &                     &              & {The authors examined knowledge transfer in \\user posts and categorized users.}                                       &                                                                                                                                                                                                              \\
\cite{samtani2016using}    &                  & \checkmark          &              & {The authors used degree and betweenness centrality \\to identify key hackers distributing keyloggers.}                & {Centrality-based methods emphasize\\network structure but fail to account \\for essential user characteristics like \\trust, reputation, and prestige, which\\are vital in underground forums.\\~}         \\
\cite{huang2021hackerrank} &                  & \checkmark          &              & {They introduced HackerRank, a tool that uses an\\~improved PageRank algorithm to identify key hackers.}               &                                                                                                                                                                                                              \\
\cite{paracha2023sus}      &                  & \checkmark          &              & {The authors introduced INSPECT, an AI-driven framework\\~that identifies key hackers using five centrality measures.} &                                                                                                                                                                                                              \\
\cite{xu2022hghan}         &                  &                     & \checkmark   & {The authors used HGHAN to develop a system for to\\identify hacker groups.}                                           & {Due to the fusion strategy HGHAN\\dilutes node features.}                                                                                                                                                   \\
\cite{zhang2018kadetector} &                  &                     & \checkmark   & The authors~introduced ActorHin2vec~identify key hackers.                                                              & {This approach overlooks important user \\attributes.}                                                                                                                                                       \\
\cite{zhang2019key}        &                  &                     & \checkmark   & {The authors propose an AHIN, improving user \\characterization in identifying key hackers.}                           & { This approach fails to capture complex\\
user attributes.}                                                                                                                  
\end{tblr}
\end{table*}


\section{System Overview}\label{approach}
\begin{figure*}
    \centering
    \includegraphics[width=1\textwidth]{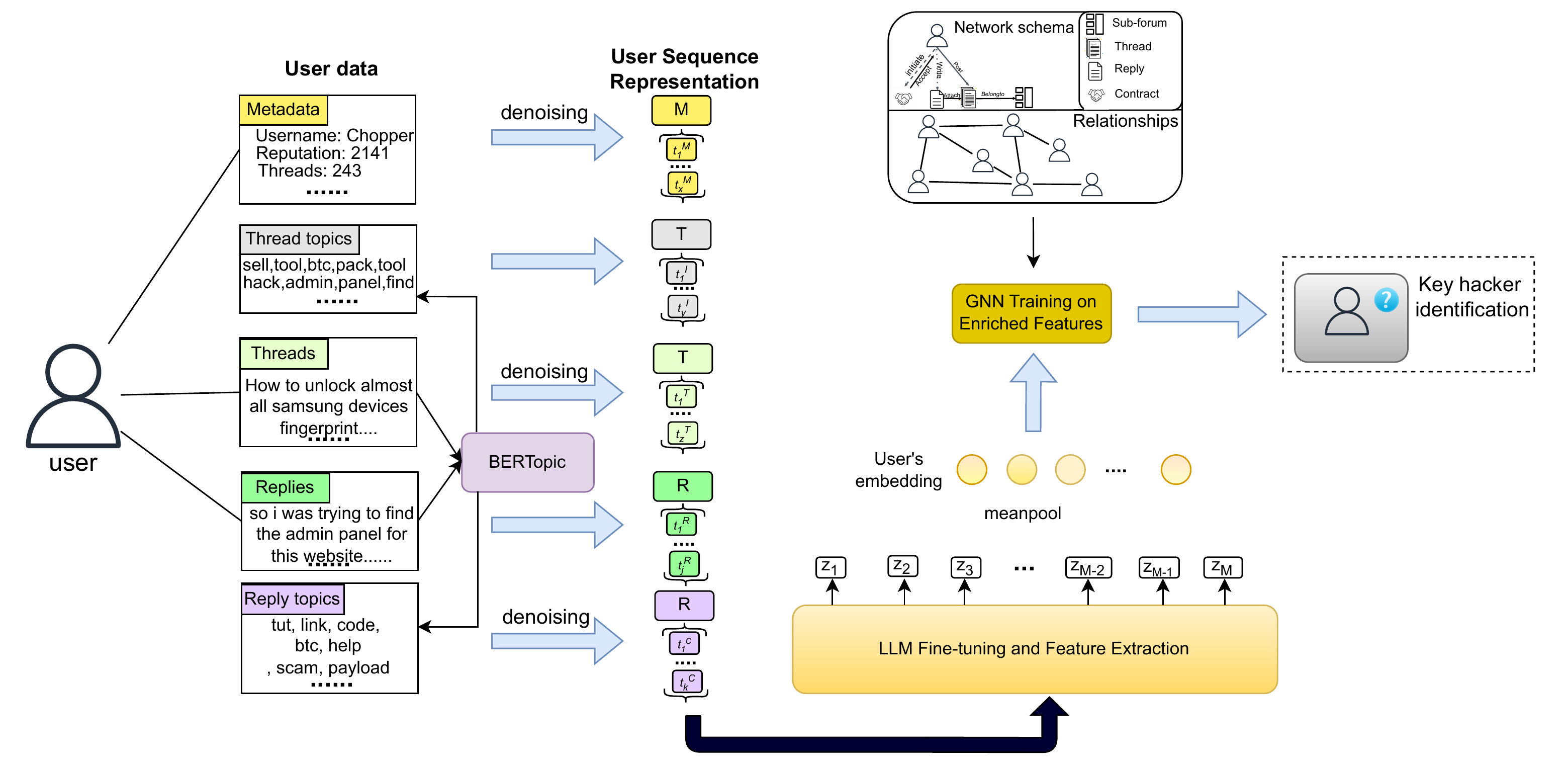}
    \caption{Overview of our proposed framework \textit{EUREKHA}.}
    \label{fig:system-overview}
\end{figure*}

In this section, we provide the system overview of \textit{EUREKHA} depicted in Fig.~\ref{fig:system-overview}. The approach consists of the following steps:
\paragraph*{\textbf{Dataset and Preprocessor}} We used CrimeBB~\cite{pastrana2018crimebb}, one of the most well-known datasets in underground forums. This dataset has been utilized in various works~\cite{paracha2023sus},~\cite{cabrero2021methodology},~\cite{pete2022postcog},~\cite{wilson2024identifying}. We focused our study on \textit{Hack-Forums}, and due to the presence of noise, we preprocessed the users' posted content to ensure it was thoroughly cleaned. In addition, the dataset is labeled to establish a reliable ground truth for our analysis. The annotation process is detailed in Section~\ref{sec:annotation}.


\paragraph*{\textbf{User Sequence Representation}} The user representation consists of two steps :
    
\paragraph*{\underline{\textit{Topics Modeling}}} Using pre-processed content from the previous module, we apply BERTopic~\cite{grootendorst2022bertopic} to generate topic distributions for each user's threads and replies, resulting in thread topics and reply topics.

\paragraph*{\underline{\textit{User as a Textual Sequence}}} Based on the generated topics from threads and replies, we now have metadata, thread topics, threads, replies, and reply topics for each user account. We evaluated various strategies for user sequence representation, exploring methods to manage long text sequences and assessing the effectiveness of these representations in enhancing model efficiency. The user sequence representations are detailed in Section~\ref{sec:representation} and their assessment is detailed in Section~\ref{user_representation}. 

\paragraph*{\textbf{LLM Fine-tuning and Feature Extraction}} To improve model stability and performance, we first perform domain adaptation. The user's text sequence is processed through the LLM, which acts as a feature extractor to produce user embeddings. These embeddings are then passed to the GNN model. 

\paragraph*{\textbf{GNN Training on Enriched Features}} The GNN utilizes user embeddings generated by LLM as input, along with a graph structure that represents various types of interactions, to effectively learn user representations. The interactions include: quoted replies, threads, and contract relationships. Based on these learned representations, the final user representations obtained from the GNN are then used to predict whether a user is a key hacker.

\section{Methodology}\label{process}
In this section, we provide details about our approach and explain how users are represented in underground forums. In addition, LLM fine-tuning and GNN training are discussed.

\subsection{Dataset and Preprocessor}
We use the latest version of the CrimeBB dataset collected on 21/06/2023. It contains data from several underground forums, with a focus on \textit{Hack-Forums}, the largest and longest-running hacking and cybercrime forum. This forum has more than 42.5 million posts and 4 million threads created by 736,000 user accounts spanning over 15 years. \textit{Hack-Forums} is structured into nine main categories: Hacking, Coding, Gaming, Technology, Market, Money, Graphics, and Common (which includes subforums for discussions on various topics, such as entertainment or politics, as well as forums for rules and suggestions). In~\cite{pastrana2018characterizing, zhang2018idetector}, the authors reported that cybercrime subforums are organized into distinct categories, such as 'Hacking' which covers techniques and tools for unauthorized access, and 'Market' where illegal goods and services are traded~\cite{samtani2015exploring}. Therefore, we focus our study on these two subforums. Table \ref{tab:hacking-market} provides key metrics for the Hacking and Market categories within the \textit{Hack-Forums} platform. The preprocessing step removes special characters and extra spaces, further cleaning the user content. It also replaces certain textual elements such as URLs, cited posts, quotations, and source code with the labels URL, CITE, QUOTE, and CODE. This enhances clarity and makes the analysis easier.

\begin{table}
\centering
\caption{Metrics for Hacking and Market categories in \textit{Hack-Forums}.}
\label{tab:hacking-market}
\begin{tblr}{
  column{2} = {c},
  hline{1-2,13} = {-}{},
}
\textbf{Metric}                   & \textbf{Value} \\
Number of Subforums               & 35             \\
Number of Users                   & 465,385        \\
Number of Posts                   & 12,630,580     \\
Number of Threads                 & 1,555,354      \\
Number of Contracts               & 348,340        \\
Average Posts per User            & 27.14          \\
Average Threads per User          & 3.34           \\
Replies per Thread                & 7.12           \\
Average Length of Posts (words)   & 29.13          \\
Average Length of Threads (words) & 5.19           \\
Language                          & EN             
\end{tblr}
\end{table}

\subsection{User Sequence Representation}
The user representation process involves two steps: first, managing large volumes of user content, and second, representing the user as a textual sequence in multiple ways, as detailed hereunder.
\subsubsection{Topics Modeling}
Topic modeling reveals the thematic structure within a text corpus by categorizing documents into distinct topics based on related words. This technique is crucial for understanding context, community types, user expertise, and content in forums~\cite{wang2023identifying},~\cite{hristova2022media},~\cite{uncovska2023rating}. Various text analysis methods, such as Latent Semantic Indexing (LSI)~\cite{rosario2000latent}, Probabilistic Latent Semantic Analysis (PLSA)~\cite{bosch2006scene}, Latent Dirichlet Allocation (LDA)~\cite{lda}, Non-Negative Matrix Factorization~\cite{lee2000algorithms}, Top2Vec~\cite{angelov2020top2vec}, and BERTopic~\cite{grootendorst2022bertopic}, are suited to different text analysis scenarios.

Unlike traditional methods that rely on word frequency and often struggle to capture complex semantic relationships and contextual nuances, BERTopic leverages advanced language understanding of BERT and c-TF-IDF~\cite{qaiser2018text} for more coherent topics. 
We applied BERTopic~\cite{grootendorst2022bertopic} to extract personalized topics from users' threads and replies, generating two distinct sets of topics for each user: thread topics and reply topics. For the first, by analyzing users' thread posts to extract recurring themes and central ideas. These thread topics represent the key subjects the user tends to initiate discussions on. For the latter, We applied the same process to users' replies. We identified distinct reply topics that reflect how users engage in discussions started by others. 

\begin{figure}
    \centering
    \includegraphics[width=1\linewidth]{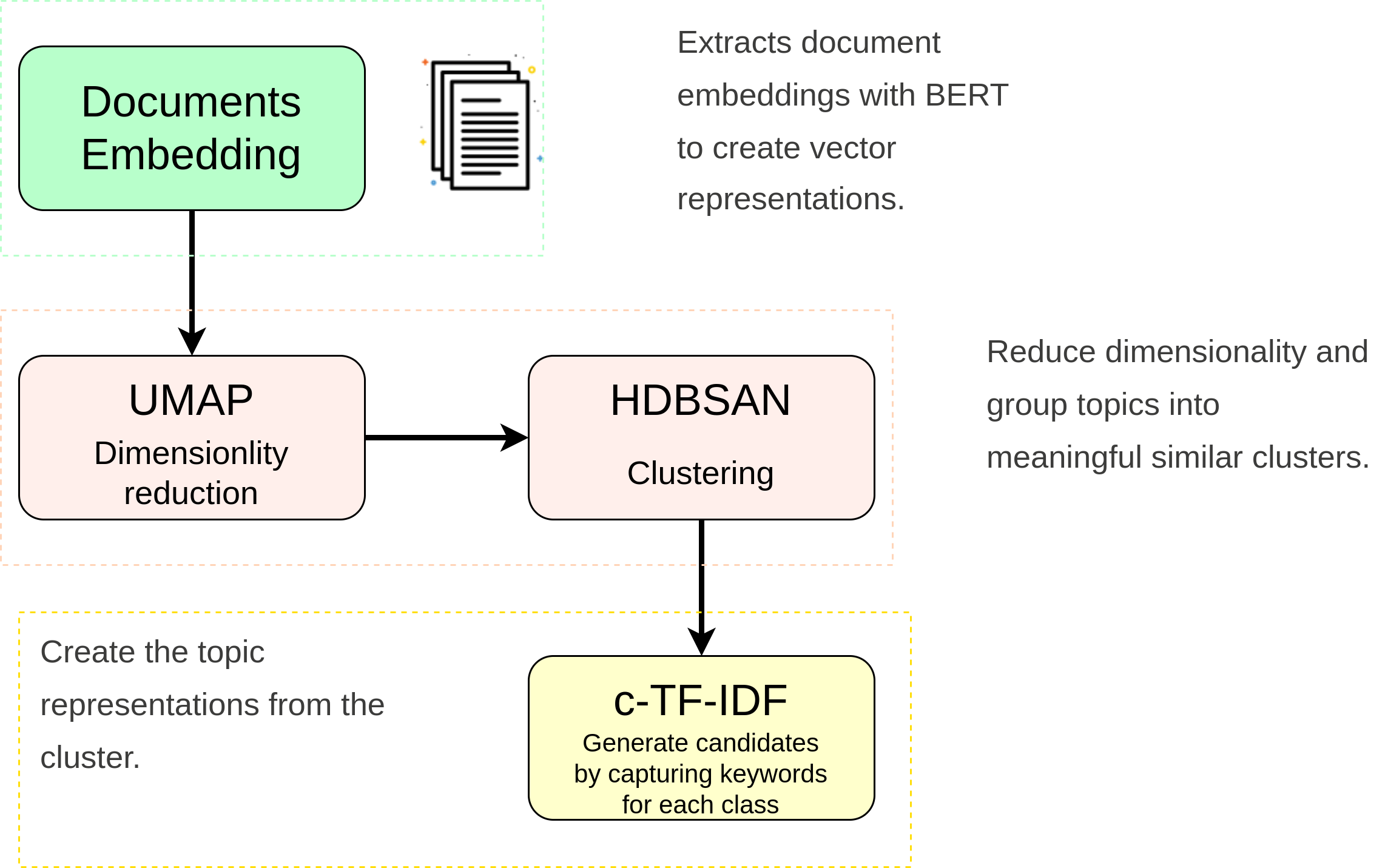}
    \caption{The workflow for modeling topics using BERTopic.}
    \label{fig:worflow-bertopic}
\end{figure}

BERTopic automatically determines the optimal number of topics, eliminating the need for manual specification. Fig.~\ref{fig:worflow-bertopic} shows the workflow of BERTopic and the details of each step are as follows: 
\begin{itemize}
    \item \textbf{Document embedding.} Each document is transformed into a vector using BERT, which captures contextual nuances and provides richer, more meaningful representations.
\item \textbf{Dimensionality reduction.} BERT embeddings are high-dimensional (often 768 dimensions). BERTopic uses UMAP~\cite{mcinnes2018umap} to reduce these dimensions while preserving key structural aspects, which prepares the data for effective clustering.

\item \textbf{Clustering.} With reduced dimensions, BERTopic uses Hierarchical Density-Based Spatial Clustering of Applications with Noise (HDBSCAN)~\cite{stewart2022implementation} for clustering, which identifies clusters of varying densities without requiring a preset number of clusters.

\item \textbf{Topic representation.} After clustering, BERTopic~\cite{grootendorst2022bertopic} uses c-TF-IDF to extract the most representative words for each topic, highlighting words that are frequent within a topic but rare across others, capturing each unique cluster.

$W_{x, c}$ represents the weight assigned to the term $x$ in the context of the cluster $c$. 
\begin{equation}
W_{x, c} = \left\|tf_{x, c}\right\| \times \log\left(1 + \frac{A}{f_x}\right)
\label{eq:weight}
\end{equation}

Where $t f_{x, c}$ represents the frequency of word $x$ within cluster $c, f_x$ indicates the frequency of word $x$ across all clusters, and $A$ is the average number of words per cluster.  
\end{itemize}

\subsubsection{User as a Textual Sequence}\label{sec:representation}
\begin{figure}
    \centering
    \includegraphics[width=1\linewidth]{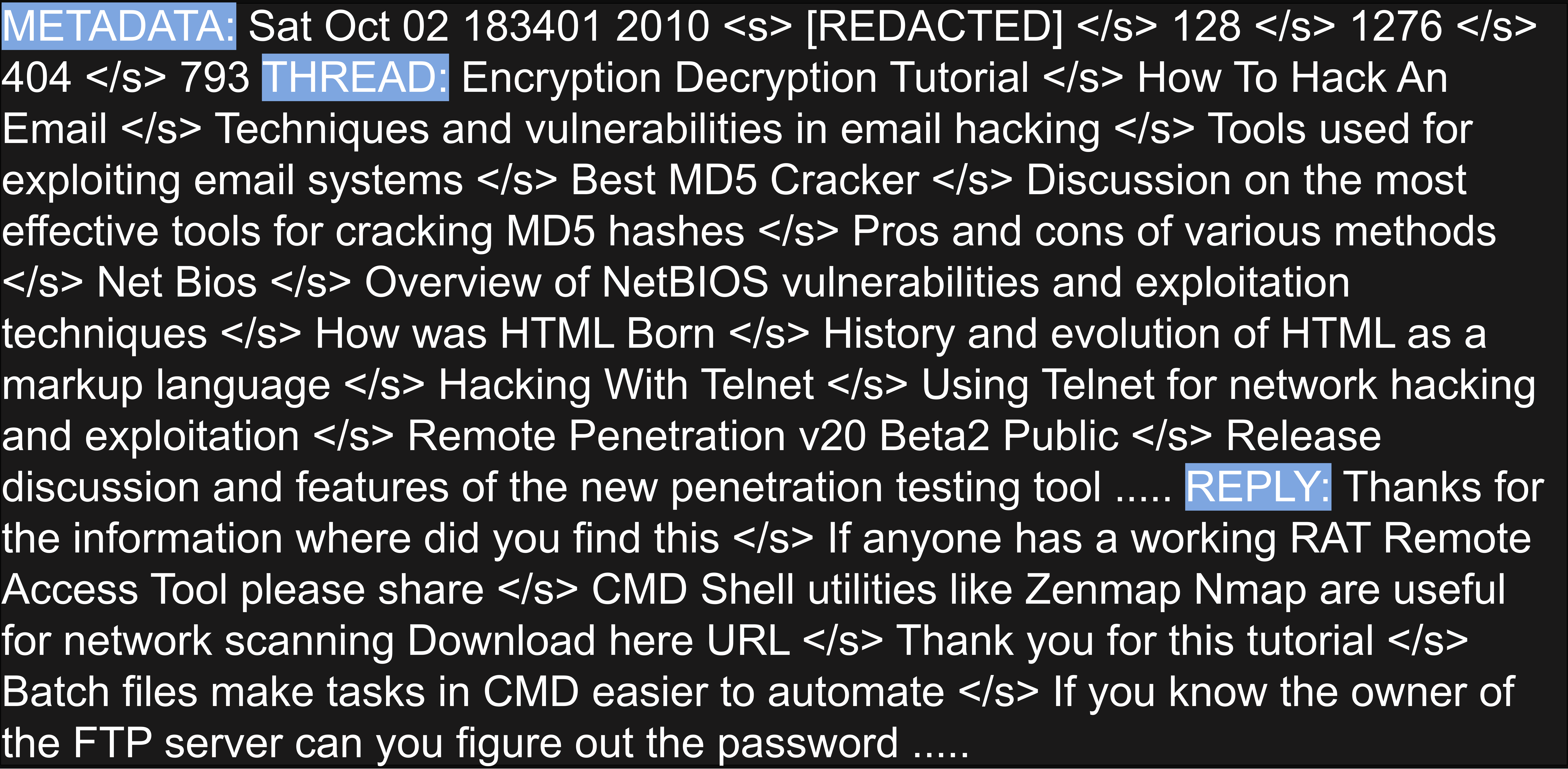}
    \caption{User sequence representation example.}
    \label{fig:example-sequence}
\end{figure}

Based on generated topics of threads and replies for each user, we now have for each account: metadata, threads, thread topics, replies and reply topics. Inspired by~\cite{lmbot}, we created a unified representation by encoding user data as structured text sequences compatible with LLM inputs. 
Specifically, for users’ metadata, \textit{EUREKHA}  extracts the corresponding information: username, thread count, post count, and reputation and organizes it in the following format:

\begin{center} [M]\{username\}[SEP]\{thread\_count\}[SEP]\{post\_count\} ..., \end{center}
Where [M] is a special token~\cite{yang2022parameter} marking the beginning of the metadata section, while [SEP] separates the metadata attributes. The same encoding procedure is applied for thread topics, threads, replies, and reply topics.  
Fig.~\ref{fig:example-sequence} shows an example of a user sequence representation.

To determine the optimal format for capturing a complete user profile, we explore different strategies to represent user sequences as detailed in Table~\ref{tab:representation_formats}:
\begin{itemize} 
\item Full Representation (R1): Combines all user metadata, full threads, and full replies into one textual sequence representing the user, without any topic summarization.
\item Reduced Threads (R2): Replaces full threads with their summarized thread topics, while keeping metadata and full replies unchanged.
\item Reduced Replies (R3): Replaces full replies with their summarized reply topics, retaining metadata and full threads.
\item Reduced Threads and Replies (R4):  Replaces both threads and replies with their corresponding topics, combining with metadata for a topic-centered representation. 
\end{itemize}
Where [M], [T], and [R] are special tokens \cite{yang2022parameter} added to the LLM's tokenizer.

\begin{table}
\centering
\caption{User Sequence Representation formats.}

\label{tab:representation_formats}
\begin{tblr}{
  row{1} = {c},
  row{5} = {c},
  cell{2}{1} = {c},
  cell{3}{1} = {c},
  cell{4}{1} = {c},
  hline{1-2,6} = {-}{},
}
\textbf{Representation} & \textbf{Format}                                                 \\
\textbf{R1}                 & {[}M] \{metadata\} {[}T] \{thread\} {[}R] \{reply\}             \\
\textbf{R2}                 & {[}M] \{metadata\} {[}T] \{thread topic\} {[}R] \{reply\}       \\
\textbf{R3}                 & {[}M] \{metadata\} {[}T] \{thread\} {[}R] \{reply topic\}       \\
\textbf{R4}                 & {[}M] \{metadata\} {[}T] \{thread topic\} {[}R] \{reply topic\} 
\end{tblr}
\end{table}

\subsection{LLM Fine-tuning and Feature Extraction}
We perform domain-adaptive fine-tuning of the LLM to improve the representation of users and increase the performance of our GNN model using user sequences and their corresponding labels. We use LLM to encode the textual sequence of the user, which can be formulated as follows: 
\begin{equation}
z_i = \frac{1}{N_i} \sum_{j=1}^{N_i} \operatorname{LLM}\left(\boldsymbol{t}_i\right)_j,
\label{eq:my_equation}
\end{equation}

Where $\boldsymbol{t}_i$ is the textual sequence of the $i$-th user, $\mathrm{LLM}(\cdot)$ denotes the large language model adopted as feature extractor, and $N_i$ is the number of tokens in $\boldsymbol{t}_{\boldsymbol{i}}$. We use mean-pooling on the LLM output to generate 768-dimensional embeddings for representing users. To make predictions, we apply an \textit{L}-layer Multilayer perceptron (MLP)~\cite{taud2018multilayer}  to reduce the dimension of $z_i$, project it into a binary classification space, and obtain the predicted logit, which is formulated as:
\begin{equation}
z_i^{(l)}=\operatorname{LeakyReLU}\left(\boldsymbol{W}^{(l)} \cdot z_i^{(l-1)}+\boldsymbol{b}^{(l)}\right),
\end{equation}

Where $z_i^{(0)}$ is the output of the LLM in Equation~(\ref{eq:my_equation}). We then apply SoftMax to the output $\operatorname{logit} z_i^{(L)}$ to obtain the prediction.
\subsection{GNN Training on Enriched Features}
For deep user representation in the underground forum, the embeddings generated by the pretrained LLM are refined by a GNN layer to model complex interactions. \textit{EUREKHA} uses GNNs to encode graph knowledge, which is expressed as:

 \begin{align} 
\begin{aligned}
& \boldsymbol{a}_u^{(l)} = \underset{v \in \mathcal{N}(u)}{\operatorname{AGGREGATE}}\left[\operatorname{MSG}^{(l)}\left(\boldsymbol{h}_u^{(l-1)}, \boldsymbol{h}_v^{(l-1)}\right)\right] \\
& \boldsymbol{h}_u^{(l)} = \operatorname{UPDATE}\left(\boldsymbol{h}_u^{(l-1)}, \boldsymbol{a}_u^{(l)}\right)
\end{aligned}
 \end{align}

Where $\operatorname{MSG}(\cdot, \cdot, \cdot)$ represents the process of passing the message from node $u$ to its neighbor $v$. The $\operatorname{AGGREGATE}(\cdot)$ function collects and combines messages from all adjacent nodes of $u$, resulting in the aggregated message $\boldsymbol{a}_u^{(l)}$. The $\operatorname{UPDATE}(\cdot, \cdot)$ function then refines the node representation $\boldsymbol{h}_u^{(l-1)}$ using $\boldsymbol{a}_u^{(l)}$, producing $\boldsymbol{h}_u^{(l)}$.

\begin{figure}
    \centering
    \includegraphics[width=1\linewidth]{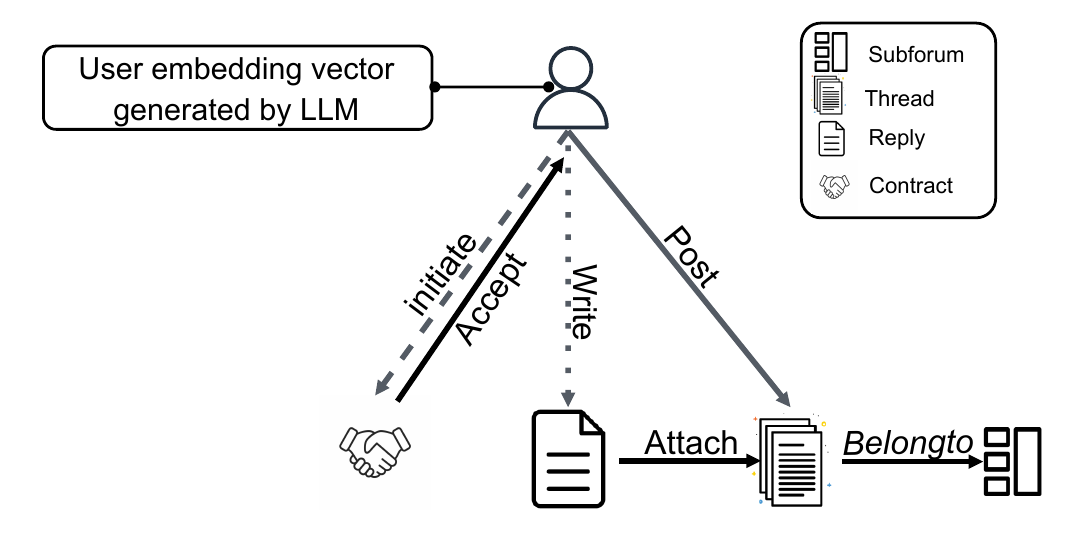}
    \caption{Network schema of \textit{Hack-Forums}.}
    \label{fig:network-schema}
\end{figure}

After processing through $L$ GNN layers, we derive the final node representations and utilize a linear transformation to compute the prediction logits for each node:
\begin{equation}
\boldsymbol{h}_i^o = \boldsymbol{W}_o \cdot \operatorname{LeakyReLU}\left(\boldsymbol{h}_i^{(L)}\right) + \boldsymbol{b}_{\boldsymbol{o}}
\end{equation}
Fig.~\ref{fig:network-schema} presents the user network schema, which allows a user to be expressively represented. Three types of relationships are captured: 
\begin{itemize}
    \item \textbf{Quoted reply relationships}: This represents direct communication where one user quotes another's reply. For instance, a \texttt{user1-quoted\_reply-user2} relationship indicates user1 quoted a reply from user2.
    \item \textbf{Thread relationships}: A \texttt{user1-thread-user2} relationship means user1 started a thread, and user2 later replied.
    \item \textbf{Contract relationships}: This denotes formal agreements between users. A \texttt{user-contract-user2} relationship indicates that user1 initiated a contract with user2, potentially involving illicit transactions.
\end{itemize}

\section{Experiments}\label{experiment}
In this section, we evaluate \textit{EUREKHA} using the \textit{Hack-Forums} dataset obtained from CrimeBB. We first outline the dataset's annotation process, followed by an assessment of the user sequence representations discussed earlier. Next, we describe the fine-tuning process and evaluate various LLM and GNN models. 

\subsection{Experiment Settings}
\subsubsection{Datasets} To evaluate our system, we use \textit{Hack-Forums}, one of the largest online underground forums. Our study focuses on two active categories, "Market" and "Hacking" which cover various cybercrime topics such as Remote Access Trojans (RATs), keyloggers, web hacking, hacking tools, malware tools, security breaches, and monetization.
We analyzed data collected from September 2007 to June 2023. Table~\ref{tab:hacking-market} summarizes the data for these categories.

\subsubsection{Annotations}\label{sec:annotation} To establish the ground truth for our study, we select a portion of users (i.e., 5,500 users) and perform a two-step labeling process to identify key hackers within \textit{Hack-Forums}.

\paragraph*{\textbf{Step 1: Automated Identification}} Based on the findings of~\cite{pastrana2018characterizing}, we used an automated approach to identify potential key hackers by analyzing metrics such as user activity, interests, reputation, and keywords in their threads and replies. This helped us generate a preliminary list of potential candidates by detecting patterns and anomalies associated with known key hackers.

\paragraph*{\underline{\textit{Public Records Analysis}}} Using the Google search engine, we searched for \textit{Hack-Forums} users arrested for cybercrimes with terms like "cybercrime" "Hack-Forums" "arrested" and "hackers". We identified some matching usernames but later discovered that some users may have changed their names or were not involved in cybercrime. 

\paragraph*{\underline{\textit{Behavioral Analysis}}} We identify users by features such as activity levels (at least 400 replies and 20 threads), engagement in selling and monetizing (examined through Market subforums), and reputation (greater than 100). Reputation reflects the peer evaluations of users.

\paragraph*{\underline{\textit{Keyword Analysis}}} We searched for threads advertising the top 300 Remote Access Trojans (RATs) from~\cite{valeros2018study} and included keywords such as: 'bypass', 'hacking', 'hacker', 'hack', 'shell', 'bomber', 'virus', 'bot', 'botnet', 'DDoS', and 'crypter'. Users who frequently use these terms are more likely to be involved in illicit activities.

Based on these criteria, we have identified 1,600 potential key hackers.

\paragraph*{\textbf{Step 2: Manual Verification}} To further refine the automated method findings, we conducted a detailed two-month manual review of potential key hackers based on their posted content. This process resulted in the final classification of 794 users as key hackers, while the remaining 4,706 users were classified as non-key hackers.

\subsubsection{Evaluation Metrics} We use accuracy and F1-score as evaluation metrics. 
\textbf{Accuracy} measures the overall correctness of the predictions, assessing how well the model classifies both key hackers and non-hackers. It calculates the proportion of correct predictions, including true positives (TP) and true negatives (TN), relative to the total number of cases evaluated.

\textbf{F1-score}, on the other hand, emphasizes the model's performance in correctly predicting key hackers. It balances precision (the proportion of correctly identified key hackers out of all users flagged as hackers) and recall (the proportion of actual key hackers correctly identified). This metric is crucial for evaluating the effectiveness of \textit{EUREKHA} in accurately identifying key hackers while managing false positives and negatives.

\subsubsection{Configurations} We conducted our experiments using a Scholar High-Performance Computing system.  The resources used in the experiments are detailed as follows: The server was running Linux (RedHat version) with 16 GB of RAM and a Tesla VOLTA 100 GPU. We use Python version 3.9.15.

\subsection{Baselines}\label{baselines}
To evaluate the performance of \textit{EUREKHA}, we compare it against several state-of-the-art baselines. The \textbf{DA} method~\cite{DA} characterizes users through content-based and structural features and employs the X-means algorithm to identify expert hackers and their areas of expertise. The \textbf{RRI} approach~\cite{RRI} measures user radicalness and associations, ranking users by influence using a customized PageRank algorithm. \textbf{KADetector}~\cite{zhang2018kadetector} utilizes a heterogeneous information network (HIN) to identify key hackers based on relationships among users, posts, and replies, without relying on node attributes. Finally, \textbf{iDetective}~\cite{zhang2019key} employs an attributed heterogeneous information network (AHIN) and meta-paths to model user interactions, enhancing the identification of key hackers.

\subsection{Evaluation of User Sequence Representation}\label{user_representation}
After generating user sequence representations, we noticed significant variations in sequence length, leading us to address the challenge of managing long sequences. LLMs such as BERT are limited to 512 tokens, and 100\% of the sequences in R1 and R2 exceeded this limit. In R3, 25\% of the sequences still surpassed the limit. With R4, where BERTopic is applied to both replies and threads, the sequences fit within the limited number of tokens.
To manage these long user sequences, we adopt the common strategies outlined in~\cite{sun2019fine}:

\paragraph*{\textbf{Truncation methods}} Each segment of the user sequence (metadata, threads, and replies) is assigned a specific token limit: metadata (34 tokens), threads (239 tokens), and replies (239 tokens). Allocating 26 tokens to user metadata covers all essential details about the user. The remaining 484 tokens are evenly split between user threads and replies. This approach ensures a balanced representation of the user's content within the token limit.

\paragraph*{\textbf{{Hierarchical methods}}} The lengthy user sequence representations are divided into segments of fewer than 512 tokens to comply with the LLM's token limit. Each segment is then input into the LLM to obtain its representation. The representation of each segment corresponds to the hidden state of the [CLS] token from the last layer. We then combine the representations of all segments using hierarchical mean pooling (Hierar. Mean), max pooling (Hierar. Max), and self-attention (Hier. Self-a.) techniques.

Table~\ref{table:sequence_handling} shows the performance of different user sequence representations based on the F1-score. The best result is in bold, and the worst one is underlined. User sequence R3 achieved the highest F1-score across all handling methods, indicating its effectiveness for representation and classification. In contrast, sequence R2 consistently recorded the lowest F1-score, especially with the Hierarchical Mean method. This suggests that R2 may contain more noise or fewer distinguishing features, making it difficult for representation models to capture meaningful patterns, which highlights the importance of preserving the thread's integrity.

The "Truncation" method is more effective than the "Hierarchical Self-Attention" method, though the latter still yields competitive results. Conversely, the "Hierarchical Mean" method performs the weakest, suggesting that averaging may dilute important information.

R1 relies on full content without BERTopic likely which leads to the inclusion of excessive, less relevant information, which affects its performance, particularly with more complex handling methods such as Hierarchical Mean.

With the R3 user sequence representation, the truncation method achieves the best performance. Therefore, we used this combination in the following experiments.

\begin{table}
\centering
\caption{Performance evaluation of various user sequence representations (User Rep.) using BERT models with different techniques for handling long sequences.}
\label{table:sequence_handling}
\begin{tblr}{
  row{odd} = {c},
  row{4} = {c},
  row{6} = {c},
  cell{1}{1} = {r=2}{},
  cell{1}{2} = {c=4}{},
  hline{1,3,7} = {-}{},
  hline{2} = {2-5}{},
}
\textbf{User Rep.} & \textbf{F1-Score}         &                              &                             &                              \\
                   & \textbf{\textbf{Trunca.}} & \textbf{\textbf{Hier. Mean}} & \textbf{\textbf{Hier. Max}} & \textbf{\textbf{Hier. Self}} \\
R1                 & 74.78                     & 61.12~                       & 65.78                       & 73.56                        \\
R2                 & 67.89                     & \uline{53.34}~               & 57.78                       & 62.45                        \\
R3                 & \textbf{84.39}            & 69.56~                       & 74.34                       & 76.89                        \\
R4                 & 72.45                     & 61.78~                       & 67.23                       & 70.78                        
\end{tblr}
\end{table}
\subsection{Effect of LLM fine-tuning}
We conducted a study to assess the impact of fine-tuning LLM, with the primary goal of determining whether fine-tuning is crucial for achieving optimal performance. 

Table~\ref{tab:effect-finetunning} shows that fine-tuning led to achieving the highest accuracy and F1-score, emphasizing the importance of fine-tuning LLMs for better performance. We enhance this performance further with hyperparameter tuning as described in the following subsection.
\begin{table}
\centering
\caption{Impact of Fine-Tuning of LLM using user sequence representation R3.}
\label{tab:effect-finetunning}
\begin{tblr}{
  cell{2}{1} = {r=4}{c},
  cell{2}{3} = {c},
  cell{2}{4} = {c},
  cell{3}{3} = {c},
  cell{3}{4} = {c},
  cell{4}{3} = {c},
  cell{4}{4} = {c},
  cell{5}{3} = {c},
  cell{5}{4} = {c},
  cell{6}{1} = {r=4}{c},
  cell{6}{3} = {c},
  cell{6}{4} = {c},
  cell{7}{3} = {c},
  cell{7}{4} = {c},
  cell{8}{3} = {c},
  cell{8}{4} = {c},
  cell{9}{3} = {c},
  cell{9}{4} = {c},
  cell{10}{1} = {r=4}{c},
  cell{10}{3} = {c},
  cell{10}{4} = {c},
  cell{11}{3} = {c},
  cell{11}{4} = {c},
  cell{12}{3} = {c},
  cell{12}{4} = {c},
  cell{13}{3} = {c},
  cell{13}{4} = {c},
  hline{1-2,6,10,14} = {-}{},
}
\textbf{Strategy}                                       & \textbf{Model} & \textbf{Accuracy} & \textbf{F1-score} \\
{Without Fine-Tuning \\(BERT)}                 & GAT   & 86.45    & 38.68    \\
                                               & GATv2 & 86.18    & 38.21    \\
                                               & GCN   & 84.81    & 35.01    \\
                                               & RGCN  & 86.09    & 40.92    \\
{Fine-Tuning\\(BERT-random hyperparameter)   } & GAT   & 88.36    & 53.28    \\
                                               & GATv2 & 91.81    & 70.58    \\
                                               & GCN   & 87.90    & 49.43    \\
                                               & RGCN  & 92.09    & 71.66    \\
{Fine-Tuning\\(BERT-tuned hyperparameter)   }   & GAT   & 95.81    & 82.45    \\
                                               & GATv2 & 95.73    & 86.11    \\
                                               & GCN   & 90.18    & 58.14    \\
                                               & RGCN  & 95.54    & 87.07    
\end{tblr}
\end{table}

\subsection{LLMs Hyperparameter Tuning}
In~\cite{bert}, it has been observed that when models are trained on smaller datasets, they become more sensitive to hyperparameter choices, which increases the risk of overfitting. To address this issue, we conducted a grid search over a set of pre-defined hyperparameters recommended by the BERT authors, including \textbf{batch size} [16, 24], \textbf{learning rate} [1e-5, 5e-5], and \textbf{epochs} [1, 5], to find the best settings. Each experiment was conducted five times with shuffled data, and the results were averaged for consistency. We evaluated four optimizers: Adam, AdamW, Adadelta, and RAdam, finding that AdamW yielded the best performance. The configuration parameters included a weight decay of 0.01, a learning rate warmup of 0.6, and a linear decay of 0.1, alongside a two-layer LLM classifier. Additionally, we applied a dropout rate of 0.1 and used leaky ReLU activation.  We fine-tuned BERT~\cite{koroteev2021bert}, RoBERTa~\cite{liu1907roberta}, ALBERT~\cite{he2020deberta}, and XLNet~\cite{yang2019xlnet} using the R3 user sequence representation, which was divided into training (60\%), validation (20\%), and testing (20\%) sets.

\begin{figure}
    \centering
    \includegraphics[width=1\linewidth]{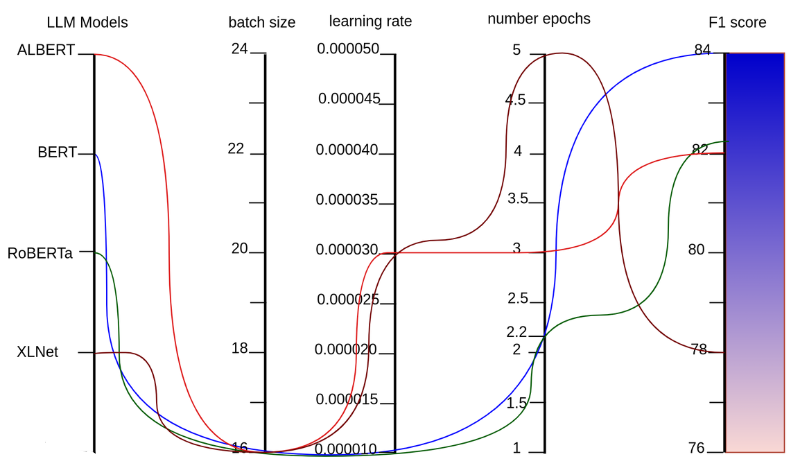}
    \caption{Best hyperparameter configurations for ALBERT, BERT, RoBERTa, and XLNet on the R3 user sequence representation. }
    \label{fig:hyper-param}
\end{figure}

\begin{figure}
    \centering
    \includegraphics[width=1\linewidth]{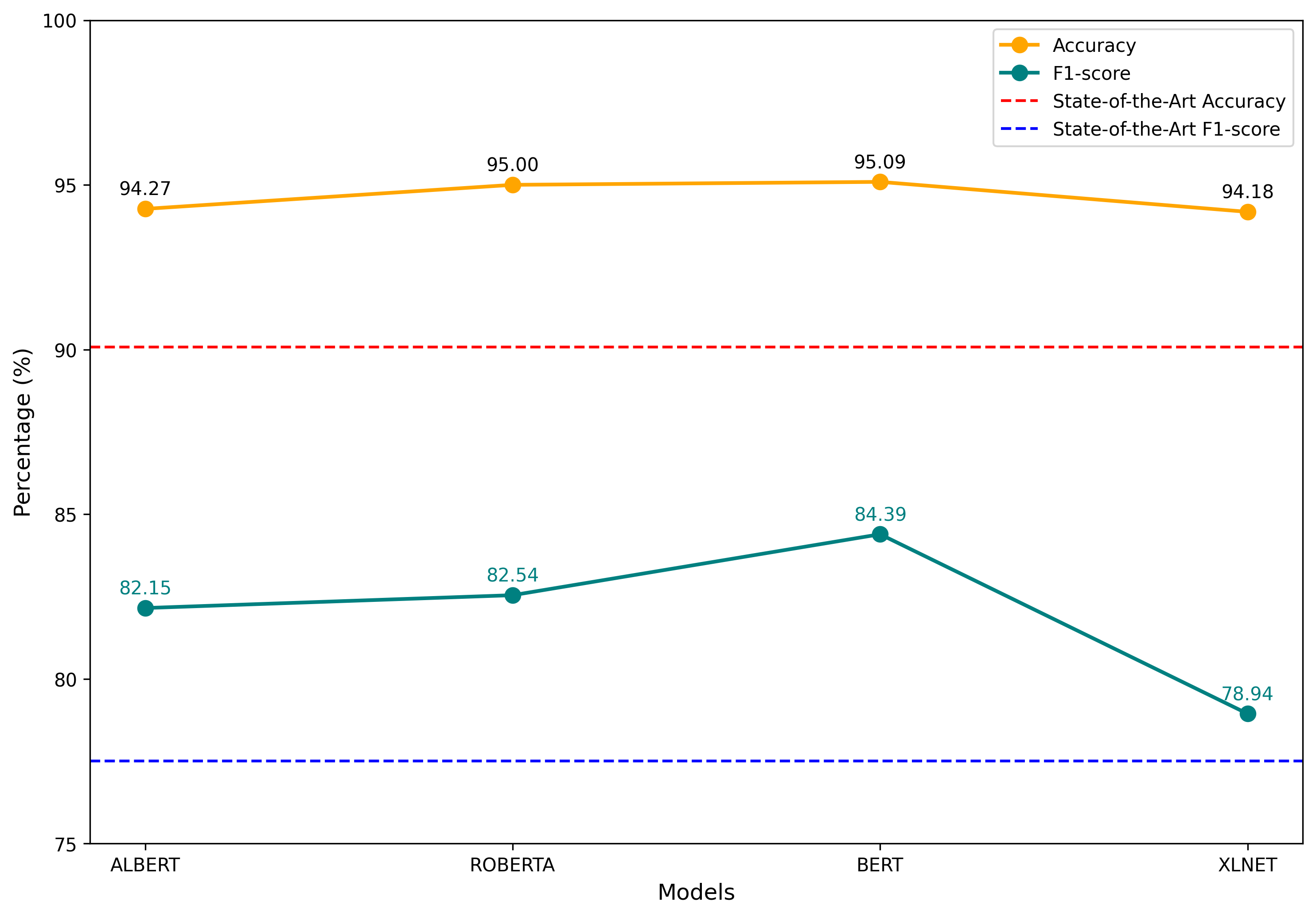}
    \caption{Performance of fine-tuned LLMs.
    }
    \label{fig:model-accuracy}
\end{figure}

Fig.~\ref{fig:hyper-param} illustrates the optimal hyperparameter combinations for each LLM model, evaluated based on the F1-score, which was prioritized due to the imbalanced nature of the dataset. Moreover, Table~\ref{tab:effect-finetunning} illustrates that LLM hyperparameter tuning achieved the best performance results, outperforming the current state-of-the-art methods, as shown in Fig.~\ref{fig:model-accuracy}.

Furthermore, Table~\ref{tab:detail-llm} presents a detailed comparison of these LLMs. Fine-tuned LLMs consistently outperform state-of-the-art GNN-based methods. BERT, in particular, leads the group with an accuracy of 95\% 

XLNet required the longest training time (170 minutes), significantly more than RoBERTa and BERT, completed in 29–30 minutes. This extended training time is primarily due to the number of epochs needed for XLNet to converge effectively. XLNet, despite its advanced capabilities and strong performance in text-rich tasks, relies on a complex permutation-based architecture.  

The strong performance of these models in text-heavy tasks, without relying on graph structures, underscores their efficiency in identifying key hackers in underground forums.


\subsection{Combination of Different LLMs and GNNs}
\begin{figure}[!h]
    \centering
    \begin{minipage}{0.30\textwidth} 
        \centering
        \includegraphics[width=\linewidth]{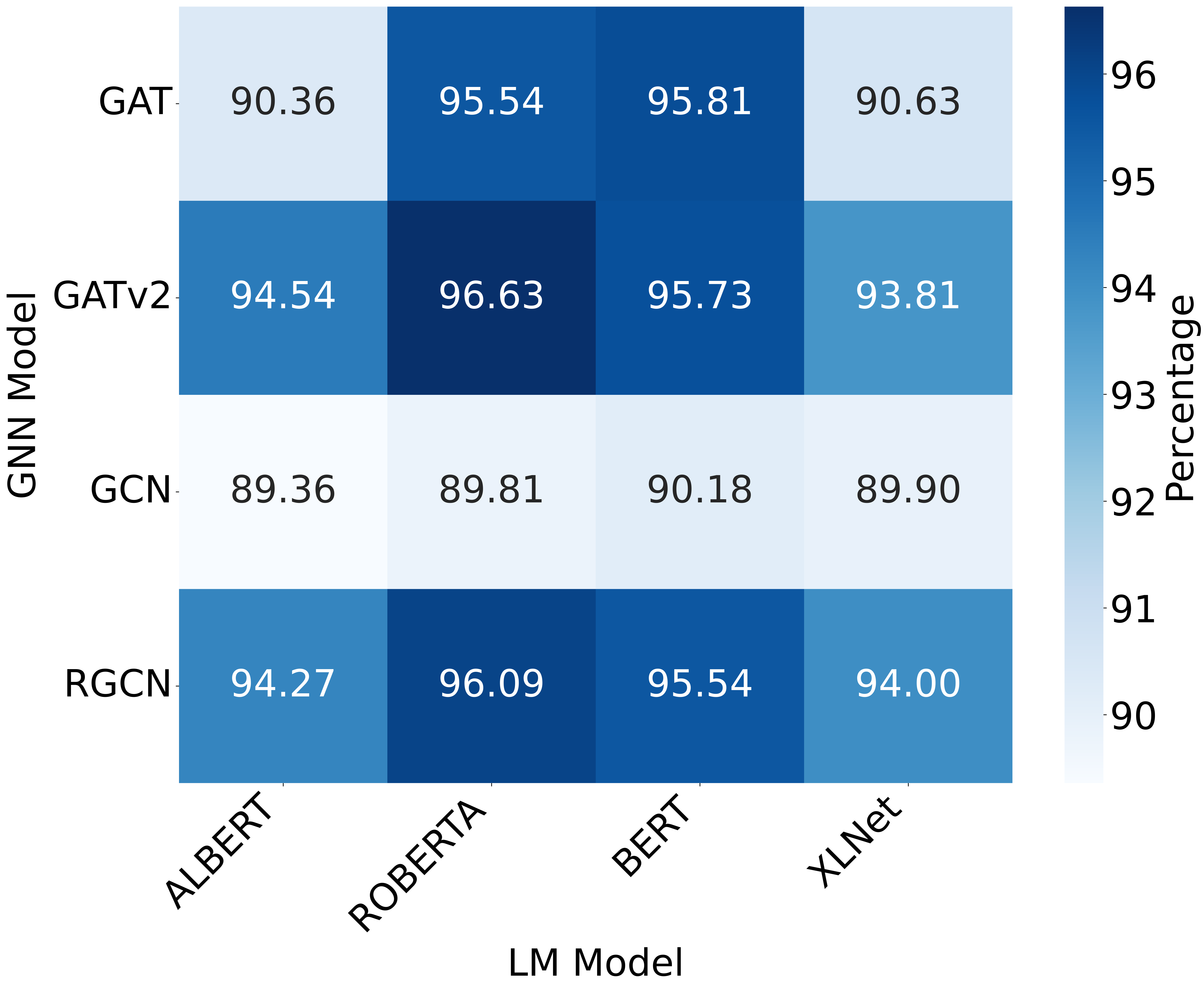} 
        \subcaption{Accuracy}
        \label{fig:figure1}
    \end{minipage}
    \hfill
    \begin{minipage}{0.30\textwidth}
        \centering
        \includegraphics[width=\linewidth]{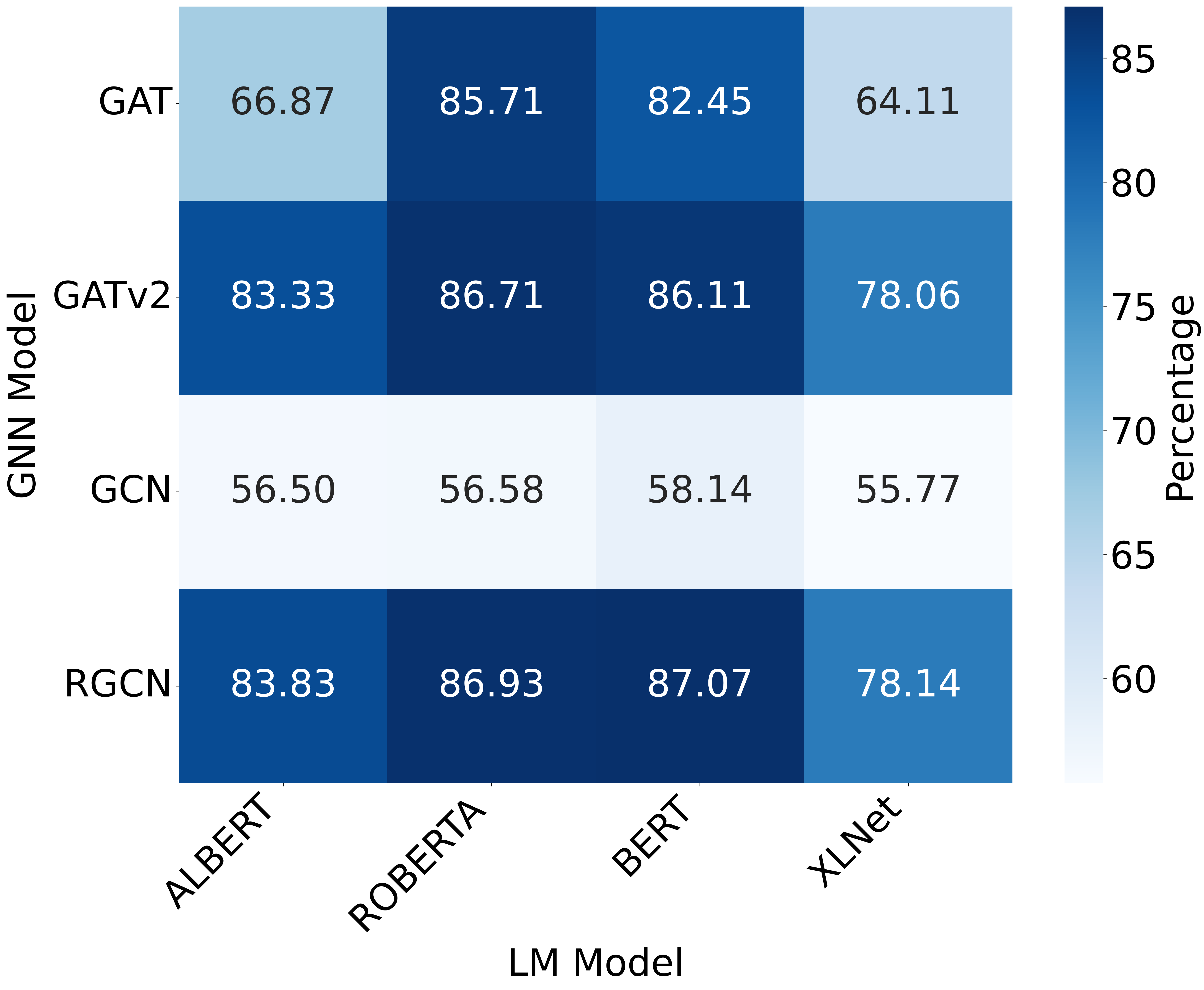} 
        \subcaption{F1-score}
        \label{fig:figure2}
    \end{minipage}
    \caption{Performance comparison under different combinations of LLMs and GNNs}
    \label{fig:combined}
\end{figure}

LLMs have varying levels of pre-existing knowledge and inductive biases. Moreover, diverse GNNs can be applied with distinct focus areas. Hence, we study the effects of combining different GNNs with LLMs to evaluate the robustness of our framework. Specifically, we explored combinations of four LLMs and four GNNs, resulting in 16 possible configurations. The LLMs selected were BERT, RoBERTa, ALBERT, and XLNet while the chosen GNNs were GCN~\cite{kipf2016semi}, RGCN~\cite{schlichtkrull2018modeling}, GAT~\cite{velivckovic2017graph}, and GATv2~\cite{brody2021attentive}. The experimental results on the R3 representation are shown in Fig.~\ref{fig:combined}. 

The differences in the GNN architectures influence significantly the performance of \textit{EUREKHA}. GATv2 and RGCN are the most effective GNNs, with the RoBERTa+GATv2 and BERT+RGCN combinations achieving the highest accuracy (96.63) and F1-score (87.07), respectively. This is likely due to their advanced attention mechanisms and relational reasoning capabilities. In contrast, the GCN underperforms, largely due to its lack of attention to node-specific relationships.

The combination of a strong LLM (e.g., RoBERTa, BERT) with a powerful GNN (e.g., GATv2, RGCN) consistently delivers the best results. This highlights the importance of selecting the right combination of the GNN and LLM to achieve optimal performance in tasks involving graph-structured data.

 The choice of LLMs also plays a crucial role in overall performance. RoBERTa and BERT  have emerged as the most adaptable and effective models across different GNN architectures. The most obvious improvement can be seen in GATv2, which achieves an F1-score of 87.07 when combined with BERT, compared to only 78.06 when it is combined with XLNet.

\

\begin{table*}
\centering
\caption{Detailed Performance Comparison of LLM Models.}
\label{tab:detail-llm}
\begin{tblr}{
  column{2} = {c},
  column{5} = {c},
  column{6} = {c},
  column{7} = {c},
  column{8} = {c},
  cell{1}{1} = {r=2}{c},
  cell{1}{2} = {c=3}{},
  cell{1}{5} = {c=2}{},
  cell{1}{7} = {c=2}{},
  cell{1}{9} = {r=2}{c},
  cell{1}{10} = {r=2}{},
  cell{1}{11} = {r=2}{},
  cell{2}{4} = {c},
  cell{3}{3} = {c},
  cell{3}{4} = {c},
  cell{3}{9} = {c},
  cell{3}{11} = {c},
  cell{4}{3} = {c},
  cell{4}{4} = {c},
  cell{4}{9} = {c},
  cell{4}{11} = {c},
  cell{5}{3} = {c},
  cell{5}{4} = {c},
  cell{5}{9} = {c},
  cell{5}{11} = {c},
  cell{6}{3} = {c},
  cell{6}{11} = {c},
  hline{1,3,7} = {-}{},
  hline{2} = {2-8}{},
}
\textbf{Models}  & \textbf{Training} &          &         & \textbf{Validation} &          & \textbf{Test}     &          & \textbf{Parameters} & \textbf{Memory}   & \textbf{Total time(min.)} \\
        & \textbf{Accuracy} & \textbf{F1-score} & \textbf{Loss}    & \textbf{Accuracy}   & \textbf{F1-score} & \textbf{Accuracy} & \textbf{F1-score} &            &          &                  \\
ALBERT  & 94.27    & 82.15    & 0.04263 & 96.00      & 86.66    & 94.27    & 82.15    & 11783682   & 11800MiB & 41               \\
BERT    & 95.09    & 84.39    & 0.05083 & 97.09      & 87.00    & 95.09    & 84.39    & 109588098  & 14718MiB & 30               \\
RoBERTa & 95.00    & 82.54    & 0.14916 & 96.90      & 86.59    & 95.00    & 82.54    & 124750722  & 14954MiB & 29               \\
XLNet   & 94.18    & 78.94    & 0.05456 & 95.36      & 81.85    & 94.18    & 78.94    & 116823426  & 12680MiB & 170              
\end{tblr}
\end{table*}

\begin{table}[!h]
\centering
\caption{Average accuracy and F1-score over five runs on the R3 user sequence representation of \textit{EUREKHA}, compared to other methods. The best results are highlighted in bold.}
\label{tab:user_representation}
\begin{tblr}{
  column{2} = {c},
  column{3} = {c},
  cell{2}{4} = {c},
  cell{3}{4} = {c},
  cell{4}{4} = {c},
  cell{5}{4} = {c},
  cell{6}{4} = {c},
  cell{7}{4} = {c},
  cell{8}{4} = {c},
  cell{9}{4} = {c},
  hline{1-2,6,8,10} = {-}{},
}
\textbf{Method}         & \textbf{Accuracy} & \textbf{F1-score} & \textbf{Time (min.)} \\
RRI~\cite{RRI}                & 0.722             & 0.479             & -                    \\
DA~\cite{DA}                & 0.772             & 0.545             & -                    \\
KADetector~\cite{zhang2018kadetector}          & 0.884             & 0.729             & -                    \\
iDetective~\cite{zhang2019key}          & 0.907             & 0.775             & 50                   \\

RoBERTa-finetune       & 0.950             & 0.825             & 29                   \\
BERT-finetune          & 0.950             & 0.843             & 30                   \\
\textit{EUREKHA} (RoBERTa+GATv2)    & 0.966             & \textbf{0.867}            & 31                   \\
\textit{EUREKHA} (BERT+RGCN) & \textbf{0.955}             & 0.870             & 32                   
\end{tblr}
\end{table}       
\subsection{Comparisons with Alternative Approaches}
We evaluate the proposed method, \textit{EUREKHA}, against the baselines outlined in Section~\ref{baselines}, employing the same number of users to ensure a consistent evaluation. The experimental results are presented in Table~\ref{tab:user_representation}. 

\textit{EUREKHA} consistently outperforms all baseline methods. By using advanced LLMs like BERT and RoBERTa, \textit{EUREKHA} effectively captures detailed semantic information and combined with the GNNs such as GATv2, which model complex user relationships the performance improves further. \textit{EUREKHA} (RoBERTa+GATv2) achieves the highest accuracy (0.966), while \textit{EUREKHA} (BERT+GATv2) achieves the best F1-score (0.870). The strength of \textit{EUREKHA} lies in the ability of LLMs to provide GNNs with enriched features, fully utilizing user-attributed features to construct higher-level semantic and structural connections between users. This deeper representation enables \textit{EUREKHA} superior performance.

The approach iDetective~\cite{zhang2019key}, which also leverages GNNs, delivers strong results by surpassing KADetector by leveraging user-attributed features. Our approach outperforms iDetective due to its reliance on shallow text embeddings, which limits its performance.

In terms of time efficiency, our approach demonstrates superior performance, completing tasks in 29 to 32 minutes, while iDetective takes 50 minutes. In ~\cite{RRI}, ~\cite{DA}, and ~\cite{zhang2018kadetector}, the authors did not assess time evaluations. Despite the increased complexity of GNNs, our models maintain reasonable runtimes of 31 minutes for RGCN and 32 minutes for GATv2.

\subsection{Implementation Details}
We implemented our proposed \textit{EUREKHA} using PyTorch~\cite{paszke2019pytorch},
scikit-learn~\cite{pedregosa2011scikit}, PyTorch Geometric~\cite{fey2019fast},  and the Transformers~\cite{wolf2020transformers} library.  To ensure reproducibility, we configured the hyperparameters for the GNN in \textit{EUREKHA} as follows: the AdamW optimizer, a learning rate of $5 \times 10^{-4}$, a dropout rate of 0.4, a total of 2 layers, a hidden size of 128, and 200 fine-tuning epochs. The code is publicly available at: \url{https://github.com/jumbo110/EUREKHA}.
\subsection{Case Studies}
For illustration purposes, we present four key hackers on \textit{Hack-Forums}. We retrieved and analyzed all threads created by these users, uncovering several findings: (1) Many users initially focused on building their reputation by providing free tutorials and tools, which they later leveraged for larger transactions. (2) As shown in Table~\ref{tab:usershighHCU}, most identified hackers specialize in specific products or services, indicating potential value for CTIs and security firms in monitoring them. The term 'rat' (Remote Access Trojan) frequently appears, highlighting key hackers' strong association with RAT coding. Other offensive tool-related terms include 'bot', 'booter', 'crypter', and 'fud' (fully undetectable). Commerce-related terms such as 'paypal', 'btc' (Bitcoin), 'free', and 'cheap' are also common. Additionally, high occurrences of 'help', 'need', and 'question' suggest users frequently seek assistance.
Studying these key hackers identified by \textit{EUREKHA} shows that mining data from underground forums can improve our understanding of the cybercrime ecosystem, helping to develop effective interventions.

\begin{table*}[!h]
\caption{Overview of 4 Key Hackers on \textit{Hack-Forums}}
\label{tab:usershighHCU}
\begin{center}
\begin{tabular}{lcccc}
\hline
\textbf{User ID} & \textbf{Activity} & \textbf{Threads} & \textbf{Posts} & \textbf{Popularity} \\
\hline

63671 & Provides hacking tutorials, technical discussions and first member of Ub3r  & 728  & 10,000+ & 1,000+ \\
71764 & Software cracking, private data transactions (142) & 44 & 7,264 & 793 \\
1462 & Cryptography, Encryption, Decryption discussions (BTC, ETH, LTC) & 265 & 6,319 & 640 \\
63398 & Program cracking, trades activation keys, release illegal keys & 692 & 8,907 & 1,246 \\
\hline
\end{tabular}
\end{center}
\begin{flushleft}
\textbf{Note:} Ub3r: A high-ranking member on \textit{Hack-Forums}.
\end{flushleft}
\end{table*}

\section{Ethical considerations}\label{ethical}
Ethical considerations are central to our research. We have signed a Non-Disclosure Agreement (NDA) with Cambridge University to access the CrimeBB dataset, which consists of publicly available content. Our use of the dataset is strictly limited to analyzing collective behaviors, without targeting any individual users. To minimize risks, we have taken additional precautions, such as omitting specific user details and the exact structure of the dataset.

\section{Limitations and Future work}\label{limitations}
While \textit{EUREKHA} has significantly advanced the identification of influential hackers within \textit{Hack-Forums}, several limitations persist in our framework. Firstly, our approach is currently limited to a single forum. Future research will expand this scope by analyzing hacker activity across multiple forums. Secondly, the ground truth used to identify key hackers is subjective and may be influenced by annotator bias. To address this, future work will explore content similarity~\cite{di2023hierarchical} to analyze a wider set of heterogeneous characteristics, including shared textual content, social connections, and temporal patterns. According to~\cite{pastrana2018characterizing}, these key hackers exhibit similar features that are crucial for their activities, such as profiles, interests, and social interactions. They frequently engage in discussions focused on hacking-related topics, emphasizing their shared context and expertise~\cite{nunes2016darknet}. It could also be interesting to explore other LLM models like T5 and GPT.

\section{Conclusion}\label{conclusion}
In this paper, we propose \textit{EUREKHA}, an advanced system designed for key hacker identification. Each user is initially represented as a textual sequence. We employ BERTopic to condense extensive posting histories into coherent topics and generate multiple representations of user sequences. LLM processes these representations, performing domain adaptation and feature extraction to select the best one. The output of the LLM is then used by the GNN as features and further establishes relationships among users, enhancing performance. Our study demonstrates that fine-tuned LLMs outperform state-of-the-art methods in identifying key hackers, with further improvements when combined with GNNs. Evaluated on the \textit{Hack-Forums} dataset, \textit{EUREKHA} achieves around 6\% and 10\% increases in accuracy and F1-score, respectively over state-of-the-art approaches, underscoring its effectiveness in key hacker identification in online underground forums.

\section{Acknowledgements}
The authors would like to express their gratitude to the Cambridge Cyber Crime Center, UK, for providing access to the CrimeBB dataset, which was used to conduct this research.

\bibliographystyle{plain}
\bibliography{ref}

\end{document}